\documentclass[aps, prx, twocolumn,longbibliography,superscriptaddress,amsmath,amssymb,floatfix]{revtex4}

\usepackage[version=4]{mhchem}
\usepackage{amssymb}
\usepackage{graphicx}
\usepackage{amsmath}
\usepackage{color}
\usepackage{mathrsfs}
\usepackage{float}
\usepackage{indentfirst}
\usepackage{mathrsfs}
\usepackage{float}
\usepackage{indentfirst}
\usepackage{textcomp}
\usepackage{comment}
\usepackage{mathtools}
\usepackage{natbib,hyperref}
\usepackage{soul}  
\usepackage{bm}

\begin{document}
\title{Disentangling Kitaev Quantum Spin Liquid}
%%%%%%%%%%%%

%%%%%%%%%%%%
\author{Xiang Li}  \altaffiliation{These authors contributed equally to this work.}
\affiliation{Key Laboratory of Artificial Structures and Quantum Control (Ministry of Education),  School of Physics and Astronomy, Shanghai Jiao Tong University, Shanghai 200240, China}

\author{Xiangjian Qian}  \altaffiliation{These authors contributed equally to this work.}
\affiliation{Tsung-Dao Lee Institute, Shanghai Jiao Tong University, Shanghai 200240, China}
\affiliation{Key Laboratory of Artificial Structures and Quantum Control (Ministry of Education),  School of Physics and Astronomy, Shanghai Jiao Tong University, Shanghai 200240, China}

\author{Mingpu Qin} \thanks{qinmingpu@sjtu.edu.cn}
\affiliation{Key Laboratory of Artificial Structures and Quantum Control (Ministry of Education),  School of Physics and Astronomy, Shanghai Jiao Tong University, Shanghai 200240, China}

\affiliation{Hefei National Laboratory, Hefei 230088, China}

\date{\today}

%%%%%%%%%%%%

%%%%%%%%%%%%
\begin{abstract}
In this work, we investigate the Kitaev honeycomb model employing the recently developed Clifford Circuits Augmented Matrix Product States (CAMPS) method. While the model in the gapped phase is known to reduce to the toric code model—whose ground state is entirely constructible from Clifford circuits—we demonstrate that the very different \emph{gapless} quantum spin liquid (QSL) phase can also be significantly disentangled with Clifford circuits. Specifically, CAMPS simulations reveal that approximately two-thirds of the entanglement entropy in the isotropic point arises from Clifford-circuit contributions, enabling dramatically more efficient computations compared to conventional matrix product state (MPS) methods. Crucially, this finding implies that the Kitaev QSL state retains significant ``Clifford-simulatable" structure, even in the gapless phase with non-abelian anyon excitations when time reversal symmetry is broken. This property not only enhances classical simulation efficiency significantly but also suggests substantial resource reduction for preparing such states on quantum devices. As an application, we leverage CAMPS to study the Kitaev-Heisenberg model and determine the most accurate phase boundary between the anti-ferromagnetic phase and the Kitaev QSL phase in the model. Our results highlight how Clifford circuits can effectively disentangle the intricate entanglement of Kitaev QSLs, opening avenues for efficiently simulating related and similar strongly correlated models.

\end{abstract}

%%%%%%%%%%%%

%%%%%%%%%%%%
\maketitle

{\em Introduction - }
\begin{figure}[t]
    \includegraphics[width=80mm]{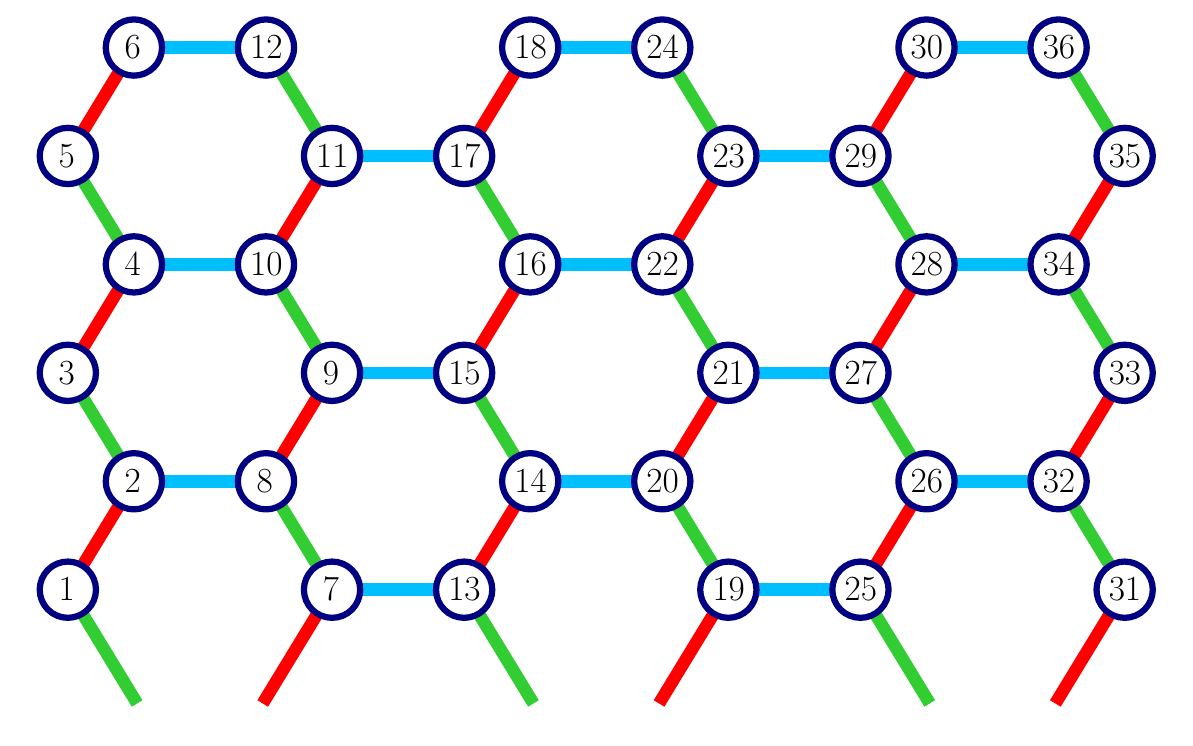}
       \caption{%Schematic representation of the honeycomb lattice and the boundary conditions employed in this work. The horizontal direction is taken to be periodic, while the vertical direction remains open, as indicated by the dashed green links. The three types of nearest-neighbor bonds characteristic of the Kitaev model are labeled as $X$, $Y$, and $Z$. To apply matrix-product-state methods, the two-dimensional lattice is mapped onto a one-dimensional chain. The orange dashed line illustrates the snake-like ordering used to construct the corresponding MPS representation, from the bottom left to the top right, which is also adopted in our CAMPS calculations.
       Schematic of the honeycomb lattice geometry and boundary conditions used in this work. Periodic boundary conditions are applied along the vertical direction, while the horizontal direction is open. The three types of Kitaev bonds ($x$ (green), $y$ (red), and $z$ (blue)) are indicated by differently colored lines. The mapping of the two-dimensional lattice ($6 \times 6$ in the plot) to a one-dimensional chain for MPS simulations is illustrated by the numbering.
       }
       \label{Kitaev_model}
\end{figure}
Strongly correlated quantum many-body systems exhibit a broad range of emergent phenomena driven by the interplay among geometry, interactions, and quantum entanglement. Despite their conceptual richness, obtaining accurate theoretical descriptions of such systems poses a great challenge due to the exponential growth of the Hilbert space and the presence of intricate long-range quantum correlations. Consequently, advanced numerical techniques have emerged as essential tools for probing the ground-state properties, low-energy excitations, and phase transitions of strongly correlated quantum materials ~\cite{PhysRevX.5.041041}. 

Quantum spin liquids (QSLs) exemplify systems in which such advanced numerical tools are particularly valuable. QSLs represent a class of quantum phases that defy classical symmetry-breaking paradigms, manifesting long-range entanglement, emergent gauge fields, and fractionalized quasi-particles \cite{RevModPhys.89.025003}. Their unconventional properties make them leading candidates for fault-tolerant quantum information processing based on topological degrees of freedom\cite{2001quant.ph..1025F,savary_quantum_2016,balents_spin_2010}.
Among theoretical models of QSLs, the Kitaev honeycomb model holds a privileged position thanks to its exact solvability and explicit realization of a $\mathbb{Z}_2$ QSL \cite{kitaev2006anyons}. 

% Among theoretical models of QSLs, the Kitaev honeycomb model occupies a distinguished position due to its exact solvability and explicit realization of a $\mathbb{Z}_2$ quantum spin liquid\cite{kitaev2006anyons}. 

Bond-directional interactions of the Kitaev type arise naturally in several materials with strong spin–orbit coupling, leading to the proposal of numerous “Kitaev material” candidates, such as Na$_2$IrO$_3$, $\alpha$-Li$_2$IrO$_3$\cite{chaloupka2013zigzag}, and $\alpha$-RuCl$_3$\cite{PhysRevB.90.041112}. While these systems often exhibit magnetic order at low temperatures, extensive experimental and theoretical works indicate that they reside near the boundary of Kitaev spin-liquid phases\cite{PhysRevB.82.064412,PhysRevB.83.220403,PhysRevLett.108.127204,PhysRevB.85.180403,PhysRevLett.109.266406,hwan_chun_direct_2015,PhysRevLett.108.127203,PhysRevB.93.195158,PhysRevB.90.041112,PhysRevLett.114.147201,PhysRevB.91.144420,PhysRevB.91.180401,PhysRevB.92.235119,PhysRevB.93.075144,banerjee_proximate_2016,do_majorana_2017}. To capture the leading perturbations relevant for these materials, the Kitaev--Heisenberg model (KHM) augments the bond-directional Kitaev interactions with isotropic Heisenberg exchange. This model displays a rich phase diagram containing gapless Kitaev QSL, and a variety of magnetically ordered states induced by the competition between Kitaev and Heisenberg terms\cite{sato_quantum_2021,zhang_variational_2021,consoli_heisenberg-kitaev_2020,janssen_heisenbergkitaev_2019,joshi_topological_2018,georgiou_spin-_2024,PhysRevLett.119.157203,chaloupka_zigzag_2013,gotfryd_phase_2017,kadosawa_phase_2023}. Resolving the ground-state properties and phase transitions within the KHM requires numerical techniques capable of handling strong anisotropy, frustration, and enhanced entanglement.

Tensor-network related methods \cite{RevModPhys.93.045003,xiang2023density}, such as the density matrix renormalization group (DMRG)\cite{PhysRevLett.69.2863} and its underlying wave-function ansatz, Matrix Product States (MPS)\cite{PhysRevLett.75.3537,RevModPhys.77.259,1992CMaPh.144..443F,10.5555/2011832.2011833,SCHOLLWOCK201196}, have become indispensable tools for accurately simulating low-dimensional quantum systems. Their effectiveness stems from the controlled truncation of entanglement and the efficient variational representation of low-energy states. However, the entanglement entropy that an MPS can represent scales logarithmically with its bond dimension, which fundamentally limits its ability to describe states with high entanglement. People also developed two-dimensional tensor-network ansatze \cite{2004cond.mat..7066V,PhysRevX.4.011025,PhysRevLett.102.180406} capable of encoding the entanglement area law for systems with local interactions. However, the prohibitively expensive scaling of computational cost with bond dimension $D$ prevents reaching very large $D$, which controls the amount of entanglement that can be handled.

These challenges have motivated the development of augmented tensor-network architectures aimed at extending the expressive power of Tensor network \cite{Qian_2023}. One particularly promising approach is the Clifford Circuits Augmented Matrix Product States (CAMPS) method\cite{PhysRevLett.133.190402}, which enhances MPS by incorporating layers of Clifford circuits as disentanglers acting on the physical degrees of freedom of MPS. Leveraging the Gottesman--Knill theorem \cite{gottesman1997stabilizer,PhysRevA.70.052328,PhysRevA.73.022334}, which ensures the efficient classical simulation of Clifford circuits, this method expands the variational manifold of MPS with mild computational overhead\cite{PhysRevLett.133.190402}. Recent studies have demonstrated that CAMPS achieves substantial accuracy improvements over conventional MPS, making it a powerful tool for tackling strongly entangled quantum systems\cite{huang_clifford_2025,qian_augmenting_2025}.

%For numerical simulations, the geometry of the honeycomb lattice and the boundary conditions play a central role in determining the accessible entanglement structure. In this work, we consider cylindrical clusters with periodic boundary conditions along the horizontal direction and open boundary conditions along the vertical direction; the three types of Kitaev bonds ($X$, $Y$, and $Z$) are shown in Fig.~\ref{Kitaev model}. Since MPS-based methods require a one-dimensional ordering of lattice sites, the two-dimensional honeycomb geometry is mapped onto a one-dimensional chain using a snake-like path, also illustrated in Fig.~\ref{Kitaev model}, indicated with a black dash line.

The Kitaev honeycomb model in the gapped phase is known to reduce to the toric code model whose ground state is then entirely constructible from Clifford circuits and is trivial to handle with CAMPS (with bond dimension $1$ \cite{gxdn-zwrw}). Previous work has demonstrated that the entanglement entropy in the gapless phase results from the sum of contributions originating from both the gauge field and the fermionic degree of freedom \cite{PhysRevLett.105.080501}. This property of the entanglement structure makes us wonder whether the gapless phase can also be significantly disentangled with Clifford circuits.

To answer this question, we calculate the ground state of the Kitaev honeycomb model at the isotropic point with CAMPS in this work. We find that for the system sizes (from $8 \times 8$ to $16 \times $16) we studied, about two-thirds of the whole entanglement entropy in the ground state can be handled by the Clifford circuits, which makes CAMPS extremely efficient to study the Kitaev honeycomb model. For example, for the $16 \times 16$ system in the isotropic point, CAMPS with bond dimension $D=20$ can give more accurate energy than ordinary MPS with bond dimension $1200$. As an application, we also leverage
CAMPS to study the Kitaev-Heisenberg model and determine the most accurate phase boundary between the anti-ferromagnetic and the Kitaev spin liquid phases in the model through reliable finite-size scaling.

%to investigate the ground state of the Kitaev model and to determine the phase-transition points of the Kitaev–Heisenberg model between the Neel phase and the gapless Kitaev spin liquid(KSL) phase. By benchmarking CAMPS against conventional MPS calculations, we demonstrate that CAMPS achieves markedly improved accuracy at fixed computational cost, particularly in strongly entangled regimes near the Kitaev points. Our results highlight the capability of CAMPS to efficiently capture the entanglement structures characteristic of Kitaev-type Hamiltonians. More broadly, this study illustrates the potential of hybrid tensor-network architectures to advance the numerical exploration of correlated quantum materials.

\begin{figure*}[t]
    \includegraphics[width=180mm]{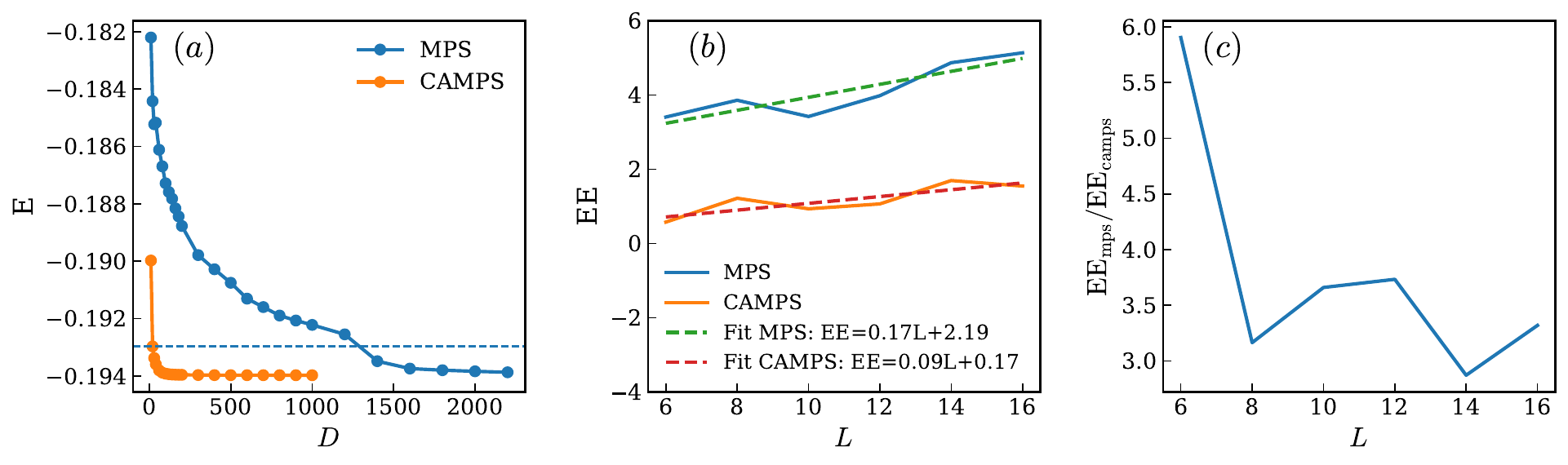}
       \caption{(a) Convergence of the ground-state energy of the Kitaev honeycomb model as a function of the bond dimension $D$ for both MPS (blue) and CAMPS (orange). The system size is $16\times16$. While the MPS energy decreases slowly and requires $D \gtrsim 2000$ to approach convergence, the CAMPS reaches a significantly lower energy already at very small bond dimensions $D \lesssim 100$. Notably, the CAMPS result at a very small bond dimension $D=20$ already achieves a lower variational energy than the MPS calculation with $D=1200$. (b) entanglement entropy (EE) as a function of system size $L$ for MPS and CAMPS. Simple linear fits (dashed lines) show that the slope of CAMPS ($0.09$) is substantially smaller than that of MPS ($0.17$).(c) Ratio of entanglement entropies, $\mathrm{EE}_{\mathrm{MPS}}/\mathrm{EE}_{\mathrm{CAMPS}}$, as a function of $L$. Across all sizes, the ratio basically remains above 2.9, confirming that MPS consistently requires more entanglement to describe the same physical state. }
       \label{Kitaev_result}
\end{figure*}

{\em Results for Kitaev model -}
%We begin by employing CAMPS on the Kitaev honeycomb model. 
The Hamiltonian of the Kitaev honeycomb model is defined as
\begin{equation}
    H = J_x\sum_{\langle ij\rangle_x} S_i^x S_j^x +J_y\sum_{\langle ij\rangle_y} S_i^y S_j^y + J_z\sum_{\langle ij\rangle_z} S_i^z S_j^z,
\end{equation}
%where $\gamma\in\{x,y,z\}$ labels the bond type, and 
where $S_i^\gamma (\gamma=x,y,z)$ are spin-$1/2$ operators on site $i$ and $J_\gamma$ is the corresponding coupling constant. The model is defined on a honeycomb lattice (see Fig.~\ref{Kitaev_model}). Each nearest-neighbor pair $\langle ij\rangle_\gamma$ interacts only through the $\gamma$-component of the spin, giving rise to a highly frustrated system. 

The kitaev honeycomb model is exactly solvable by representing the spin degree of freedom as Majorana degree of freedom \cite{kitaev2006anyons,PhysRevLett.98.087204}. The Hamiltonian is then reduced to non-interacting fermions coupled to $Z_2$ gauge fields. The phase diagram of the Kitaev honeycomb model consists of two different phases. One is the gapped phase, which carries Abelian anyon excitations, i.e., the so-called A phase. In this phase, the low-energy physics of the Kitaev model can be described by the Toric code model \cite{kitaev2006anyons}. The other one is the gapless phase, which carries non-Abelian anyon excitations if the time reversal symmetry is broken with a small magnetic field, i.e., the so-called B phase. In this work, we focused on the isotropic point (i.e., $J_x= J_y=J_z=1$), which lies in the center of the gapless B phase. 

%In this work, we consider cylindrical clusters with periodic boundary conditions along the horizontal direction and open boundary conditions along the vertical direction as shwon in Fig.~\ref{Kitaev model}. The three types of Kitaev bonds ($X$, $Y$, and $Z$) are shown in Fig.~\ref{Kitaev model} with different colors. Since MPS-based methods require a one-dimensional ordering of lattice sites, the two-dimensional honeycomb geometry is mapped onto a one-dimensional chain using a snake-like path, also illustrated in Fig.~\ref{Kitaev model}, indicated with a black dash line.
In this work, we consider cylindrical clusters with periodic boundary conditions along the vertical direction and open boundary conditions along the horizontal direction, as illustrated in Fig.~\ref{Kitaev_model}. The three types of Kitaev bonds—$x$ (green), $y$ (red), and $z$ (blue)—are depicted with distinct colors in Fig.~\ref{Kitaev_model}. To facilitate MPS-based simulations, which require a one-dimensional ordering of lattice sites, the two-dimensional honeycomb lattice is mapped onto a one-dimensional chain using a snake-like path, as indicated by the numbering in Fig.~\ref{Kitaev_model} for a $6 \times 6$ lattice.

Fig.~\ref{Kitaev_result}(a) shows the convergence of the ground-state energy as a function of the bond dimension $D$ for both MPS (DMRG) and CAMPS for system with size $16 \times 16$. Results for other system sizes can be found in the supplementary materials. To avoid the trapping of the calculation in a meta-stable state \cite{PhysRevLett.119.157203,PhysRevB.104.L020409}, we apply a small magnetic field (with strength $h=0.01$) in the [111] direction to break the conservation of the plaquette operator \cite{kitaev2006anyons}. As we can see in Fig.~\ref{Kitaev_result}(a), MPS converges slowly, requiring bond dimensions exceeding $D = 2000$ to achieve accurate results. In contrast, CAMPS shows rapid convergence with $D$, providing significantly improved ground state energy at much smaller bond dimensions ($D < 100$). Notably, the CAMPS energy at $D = 20$ already surpasses the accuracy of the MPS energy at $D = 1200$. Given the $D^3$ scaling of the cost in both methods, this result demonstrates that CAMPS is extremely more efficient than traditional MPS in handling the Kitaev honeycomb model. 
%underscoring the effectiveness of the Clifford layer in reducing the entanglement and reducing the bond dimension required for the MPS component.

To quantify the entanglement reduction achieved by CAMPS, Fig.~\ref{Kitaev_result}(b) shows the half-system entanglement entropy (EE) for system sizes ranging from $L=6$ to $L=16$. Notice that the EE for CAMPS here is actually the Non-stabilizerness Entanglement Entropy introduced in \cite{gxdn-zwrw}, i.e., the entanglement entropy which can't be reduced by Clifford circuits. For MPS, the EE grows approximately linearly with system size. A simple linear fit gives a slope of $0.17$.
In contrast, CAMPS exhibits a significantly reduced EE and a simple linear fit gives a slope of only $0.09$, demonstrating that the Clifford circuit effectively disentangles the system, thereby reducing the entanglement burden on the MPS representation.

The ratio $\mathrm{EE}_{\mathrm{MPS}} / \mathrm{EE}_{\mathrm{CAMPS}}$, shown in Fig.~\ref{Kitaev_result}(c), remains above $2.9$ across all studied system sizes, confirming that CAMPS consistently requires substantially less entanglement capacity for the MPS part to describe the same physical state. This reduction directly correlates with the faster variational convergence of CAMPS results observed in Fig.~\ref{Kitaev_result}(a), indicating that the Clifford circuit augmentation fundamentally reshapes the effective entanglement landscape of the Kitaev ground state.

These results provide an answer to the question in the introduction and demonstrate that the gapless QSL phase for the Kitaev honeycomb model can also be
significantly disentangled using Clifford circuits, although the gapless and gapped phases are quite different in nature and they are connected by a quantum phase transition. 

% The substantial entanglement reduction achieved by CAMPS at fixed bond dimension suggests that the approach is well-suited for studying more complex frustrated spin systems. We therefore apply both MPS and CAMPS to determine the phase transition in the Kitaev--Heisenberg model. Across the parameter regimes investigated, CAMPS consistently reaches converged results with bond dimensions a factor of three to five smaller than those required by conventional MPS. This improvement is especially pronounced near the Kitaev spin-liquid regime, where entanglement typically grows rapidly and limits the efficiency of standard tensor-network techniques.

% Overall, these results demonstrate that augmenting tensor-network states with Clifford circuits yields a highly efficient representation for gauge-structured quantum spin liquids. For the Kitaev model, CAMPS achieves faster energy convergence and significantly reduced entanglement growth, and these advantages translate directly into more reliable and computationally efficient determination of phase boundaries in extended models such as the Kitaev--Heisenberg system.

\begin{figure}[t]
    \includegraphics[width=90mm]{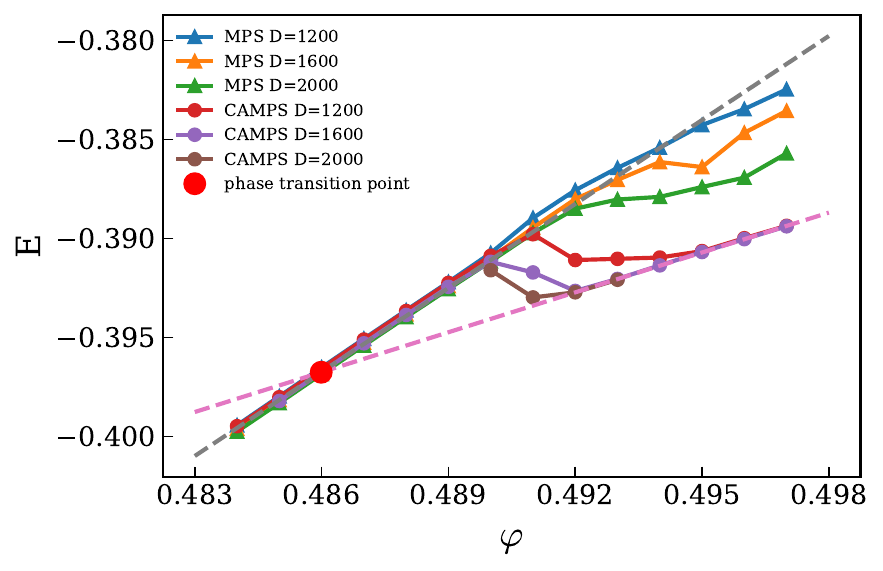}
        \caption{Ground-state energy of the Kitaev--Heisenberg model on the $16\times16$ lattice as a function of the parameter $\varphi$ (in the unit of $\pi$). The MPS data are indicated by triangles, whereas the CAMPS data are indicated by circles. The CAMPS data display significantly improvement over MPS. The dashed magenta lines represent a linear extrapolation based on the CAMPS energies. The intersection between the two lines identifies the phase-transition point, marked by the red dot. The transition point of the $16\times16$ lattice is $\varphi_c=0.48500(7)$}
        \label{Kitaev_Heisenberg_result}
\end{figure}

\begin{figure}[t]
    \includegraphics[width=90mm]{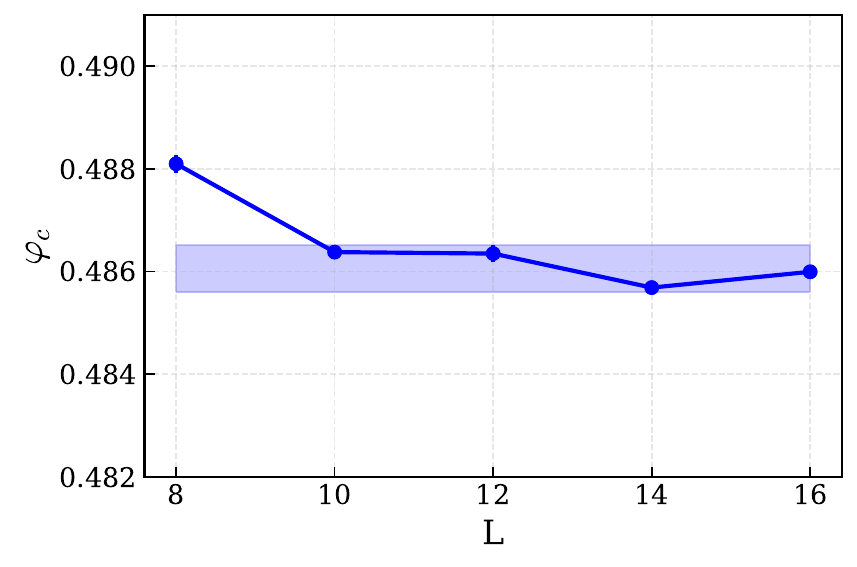}
        \caption{Finite-size estimates of the phase-transition point $\varphi_c$ (in the unite of $\pi$) obtained from CAMPS calculations for system sizes $L = 8, 10, 12, 14,$ and $16$ ($L=16$ results are shown in Fig.~\ref{Kitaev_Heisenberg_result} and other results can be found in the supplementary materials) The extracted transition points exhibit weak size dependence and converge rapidly as $L$ increases. The shaded band represents the final uncertainty range of the transition point, considering both the spread of the finite-size data and the extrapolation uncertainty.} \label{phase_transition_point}
\end{figure}

{\em Results for Kitaev-Heisenberg model -}
After demonstrating that the Kitaev honeycomb model can be efficiently handled by the CAMPS method, we also study the Kitaev--Heisenberg model on the honeycomb lattice with CAMPS. The kitaev-Heisenberg model is believed to be related to several Kitaev materials \cite{2025arXiv250105608M}. The Kitaev-Heisenberg Hamiltonian is given by
\begin{equation}
    H = \sum_{\langle i,j\rangle_\gamma} K_\gamma S_i^\gamma S_j^\gamma
        +J\sum_{\langle ij\rangle} S_i \cdot S_j, 
\end{equation}
where the first term represents the bond-dependent Kitaev interaction, coupling only the $\gamma=x,y,z$ spin components on corresponding $\gamma$-type bonds, while the second term is the SU(2)-symmetric Heisenberg exchange. Following the standard parametrization $J=\cos\varphi$ and $K_\gamma = K = 2\sin\varphi$, the pure Kitaev point is located at $\varphi=\pi/2$. 
%The pure model hosts two types of Kitaev quantum spin liquids (KSLs): a gapped $\mathbb{Z}_2$ Abelian phase (A phase) and a gapless phase with dispersing Majorana fermions and gapped flux excitations (B phase). 
Numerical studies of the Kitaev-Heisenberg phase diagram show that the gapless B-phase KSL persists for small $J$, while larger Heisenberg couplings induce several magnetically ordered phases \cite{PhysRevLett.119.157203}. The competition between anisotropic Kitaev interactions and isotropic Heisenberg exchange generates extra frustration in this model.

In our study, we use the same cylindrical lattice geometry for the Kitaev-Heisenberg model as shown in Fig.~\ref{Kitaev_model}. We focus on the phase transition between the anti-ferromagnetic phase and the Kitaev spin liquid phase in the phase diagram. Fig.~\ref{Kitaev_Heisenberg_result} shows the ground-state energy as a function of $\varphi$ for both MPS and CAMPS with different bond dimensions for the $16 \times 16$ system. The MPS results (triangle) display a pronounced and irregular drift with increasing bond dimension $D$, indicating that standard MPS (DMRG) method faces substantial difficulty converging the region near the phase boundary where entanglement is enhanced. Even at $D=2000$, the MPS optimization frequently becomes trapped in metastable, higher-energy states, producing inconsistent energy estimates and preventing reliable extrapolation. In contrast, the CAMPS results (circle) remain remarkably stable across different bond dimensions: the $D=1200$ CAMPS data already achieve lower and more consistent energies than the $D=2000$ MPS results.

%The phase-transition point can be extracted using simple linear extrapolations of the CAMPS energies in the two competing phases. The intersection of these extrapolated branches yields an reliable estimation of the critical value of \C{$\varphi_c$}, marked by the red point in Fig.~\ref{Kitaev Heisenberg result.png}.

It is known that the phase transition between the anti-ferromagnetic phase and the Kitaev QSL phase is a first order type \cite{PhysRevLett.119.157203}, so we locate the critical point from an energy-crossing analysis. We perform independent linear fits to the CAMPS ground-state energies on the two sides of the transition point and take the intersection of the fitted branches as the estimate of $\varphi_c$ (red point in Fig.~\ref{Kitaev_Heisenberg_result}).

To determine the phase transition point in the thermodynamic limit, we performed simulations for system with lengths $L=8, 10, 12, 14,$ and $16$ (the detailed results can be found in the supplementary materials). The corresponding estimates of the critical parameter $\varphi$ are shown in Fig.~\ref{phase_transition_point}. While a noticeable shift is observed between $L=8$ and $L=10$, the transition point stabilizes for larger system sizes, showing only minimal dependence on $L$. This indicates that the finite-size effects become small as $L$ increases.
The shaded region in Fig.~\ref{phase_transition_point} represents the final uncertainty range, which accounts for both the finite-size effects and the residual errors from the linear energy extrapolations. By combining these contributions, we estimate the phase-transition point to be $\varphi_c = 0.4861(5)$. For comparison, Table~\ref{tab:KH phase transition point} summarizes our result alongside those obtained from other numerical methods in the literature. All the results for the transition point are consistent in some sense, but our results is the one without systematic bias. Therefore the CAMPS result provides the best estimation for the transition point for the anti-ferromagnetic to Kitaev spin liquid phase transition in the Kitaeve-Heisenberg model.    

%The close agreement with prior studies highlights the accuracy and reliability of the CAMPS approach in determining critical points in frustrated spin systems.

% To extract the phase-transition point, we perform independent linear fits to the CA-MPS energies on both sides of the transition. In practice, the data naturally separate into two nearly linear branches corresponding to the two competing phases. We fit each branch and determine the transition point from the crossing of the two fitted lines.  This procedure yields a well-resolved transition value of $\varphi$, marked by the red point in Fig.~\ref{Kitaev Heisenberg result.png}.

% To evaluate finite-size effects, we repeat the analysis for systems of lengths $L=8,10,12,14,$ and $16$. The resulting estimates for the critical parameter $\varphi$ are shown in Fig.~\ref{phase transition point.png}. After a modest shift between $L=8$ and $L=10$, the transition point becomes essentially independent of system size. The shaded region in Fig.~\ref{phase transition point.png} represents our final uncertainty estimate, obtained from the spread in finite-size data and the residual errors of the linear energy extrapolations. Combining these contributions, we determine the phase-transition point to be $\varphi_c = 0.486057 \pm 0.000459$. A comparison of our results with those obtained using other methods is summarized in Table~\ref{tab:KH phase transition point}.

\begin{table}[t]
\centering
\begin{tabular}{lcccccc}
\hline\hline
  & ED\cite{PhysRevLett.110.097204} & fRG\cite{PhysRevB.106.174416} & CCM\cite{PhysRevResearch.6.033168} & iDMRG\cite{PhysRevLett.119.157203} &iPEPS\cite{PhysRevB.90.195102} & CAMPS \\
\hline
$\varphi_c/\pi$      & 0.489 & 0.495 & 0.472 & 0.487 & 0.489 & 0.4861(5)\\

\hline\hline
\end{tabular}
\caption{Comparison of the anti-ferromagnetic--KSL phase-transition point $\varphi_c$ obtained from different numerical methods in the literature. Exact diagnalization (ED) is for a 24-site cluster under periodic boundary conditions\cite{PhysRevLett.110.097204}. Functional renormalization group (fRG) result corresponds to a finite-size cluster containing 166 lattice sites\cite{PhysRevB.106.174416}. Coupled cluster method (CCM) result is for system in the thermodynamic limit\cite{PhysRevResearch.6.033168}. Infinite density matrix renormalization group (iDMRG) result is for infinitely long cylinder with circumference 6\cite{PhysRevLett.119.157203}. Infinite projected entangled-pair states (iPEPS) is result obtained with bond dimension $D=6$\cite{PhysRevB.90.195102}.}

\label{tab:KH phase transition point}
\end{table}

{\em Conclusion and Perspective --}
In summary, we show that the gapless quantum spin liquid phase with non-abelian anyon excitations in the Kitaev honeycomb model can be
significantly disentangled with Clifford circuits. For the system size (from $8 \times 8$ to $16 \times 16$) we studied, about two-thirds of the total entanglement entropy in the ground state of the isotropic Kitaev honeycomb model can be handled by Clifford circuits. This property not only significantly enhances classical simulation efficiency with CAMPS but also suggests substantial resource reductions for preparing such states on quantum devices. As an application, we employ CAMPS to study the Kitaev-Heisenberg model and determine the most accurate phase boundary between the anti-ferromagnetic phase and the Kitaev quantum spin liquid phase in the model, through reliable finite size scaling. Our results highlight how Clifford circuits can effectively disentangle the intricate entanglement of Kitaev QSLs, opening avenues for efficiently simulating similar strongly correlated systems that admit a description of (free) fermions coupled to a gauge field. The distinctive entanglement properties of the gapless spin phase in the Kitaev honeycomb model also indicate that generalized CAMPS methods \cite{PhysRevLett.134.150404,qian_augmenting_2025} could likewise be highly effective for the simulations of the thermodynamic and dynamic properties of Kitaev-related models, which is crucial for establishing the model Hamiltonian for Kitaev materials. We leave these investigations for future work.   

%In summary, we have shown that the Clifford circuits augmented MPS framework provides a highly efficient representation of ground states in Kitaev-type models, achieving lower energies and substantially reduced entanglement compared with conventional MPS at the same computational cost. By leveraging the disentangling power of Clifford circuits, CAMPS effectively mitigates the entanglement bottleneck that limits the scalability of standard DMRG methods. This capability is particularly advantageous for systems with emergent gauge structures or long-range entanglement, as demonstrated in our benchmark on the Kitaev model.
%Applying CAMPS to the Kitaev--Heisenberg model, we obtained a precise estimate of the N\'eel–KSL phase-transition point, $\varphi_{c}=0.486057\pm0.000459$, by conducting simulations on large system sizes. The enhanced numerical stability and rapid convergence of CAMPS allowed for reliable extrapolation of ground-state energies, even in the vicinity of the critical region where conventional MPS often struggles. These results underscore the potential of CAMPS to provide deeper insights into the phase diagrams of frustrated quantum systems.
%Looking ahead, the CAMPS framework opens up exciting opportunities for exploring a wide range of strongly correlated systems, including higher-dimensional lattices, topologically ordered phases, and systems with non-Abelian symmetries.

%%%%%%%%%%%%

\begin{acknowledgments}
\textbf{Acknowledgments:} 
We thank Hong Yao for bringing Ref.~\cite{PhysRevLett.105.080501} to our attention, which inspired us to study the Kitaev model with CAMPS during a visit to the Institute for Advanced Study at Tsinghua University, hosted by Yingfei Gu.
The calculation in this work is performed using TensorKit \cite{devos2025tensorkitjljuliapackagelargescale}. The numerical calculations were run on the Siyuan-1 high-performance computing cluster provided by the Center for High Performance Computing at Shanghai Jiao Tong University. MQ acknowledges the support from the National Key Research and Development Program of MOST of China (2022YFA1405400), the National Natural Science Foundation of China (Grant No. 12274290 and No. 12522406), and the Innovation Program for Quantum Science and Technology (2021ZD0301902).
%%%%%%%%%%%%
\end{acknowledgments}

%%%%%%%%%%%%Refs
\bibliography{main}
%%%%%%%%%%%%

\end{document}

\newcommand {\Y}{\textcolor {blue}}

\title{Supplementary Materials for ``Disentangling the Kitaev Quantum Spin Liquid''}
%%%%%%%%%%%%

%%%%%%%%%%%%
\author{Xiang Li}  \altaffiliation{These authors contributed equally to this work.}
\affiliation{Key Laboratory of Artificial Structures and Quantum Control (Ministry of Education),  School of Physics and Astronomy, Shanghai Jiao Tong University, Shanghai 200240, China}

\author{Xiangjian Qian}  \altaffiliation{These authors contributed equally to this work.}
\affiliation{Tsung-Dao Lee Institute, Shanghai Jiao Tong University, Shanghai 200240, China}
\affiliation{Key Laboratory of Artificial Structures and Quantum Control (Ministry of Education),  School of Physics and Astronomy, Shanghai Jiao Tong University, Shanghai 200240, China}

\author{Mingpu Qin} \thanks{qinmingpu@sjtu.edu.cn}
\affiliation{Key Laboratory of Artificial Structures and Quantum Control (Ministry of Education),  School of Physics and Astronomy, Shanghai Jiao Tong University, Shanghai 200240, China}

\affiliation{Hefei National Laboratory, Hefei 230088, China}

\date{\today}

\maketitle

\tableofcontents

%\appendix
\section{Results for different system sizes of the Kitaev honeycomb model}

In the main text, we presented results for the $16\times16$ system in the Kitaev honeycomb model. Here, we provide additional data for other system sizes ($8\times8$, $10\times10$, $12\times12$, and $14\times14$) to further illustrate the performance of CAMPS compared to conventional MPS.

\begin{figure*}[t]
    \includegraphics[width=80mm]{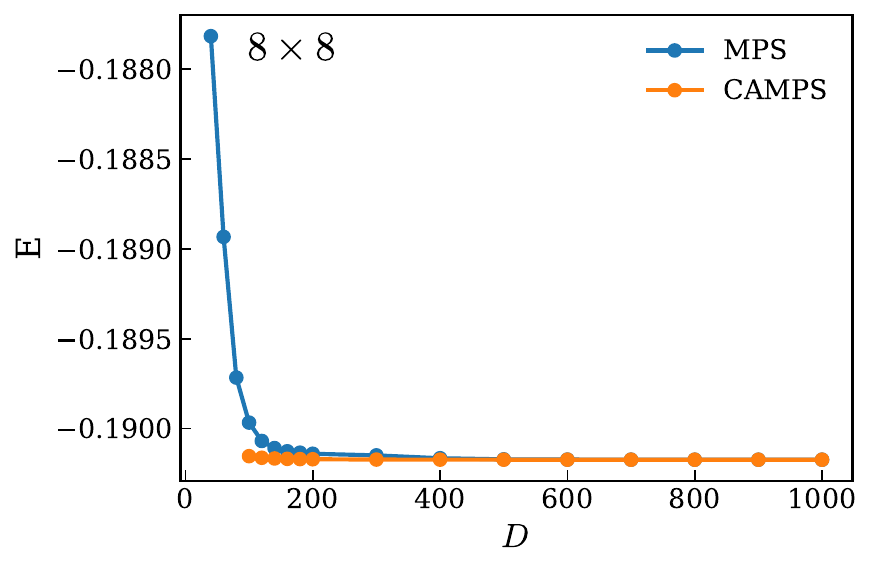}
    \includegraphics[width=80mm]{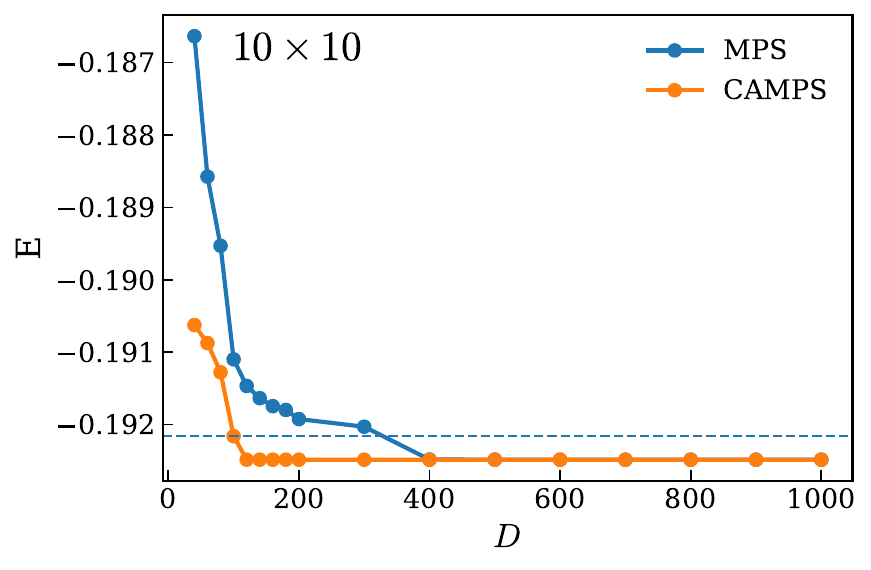}
    \includegraphics[width=80mm]{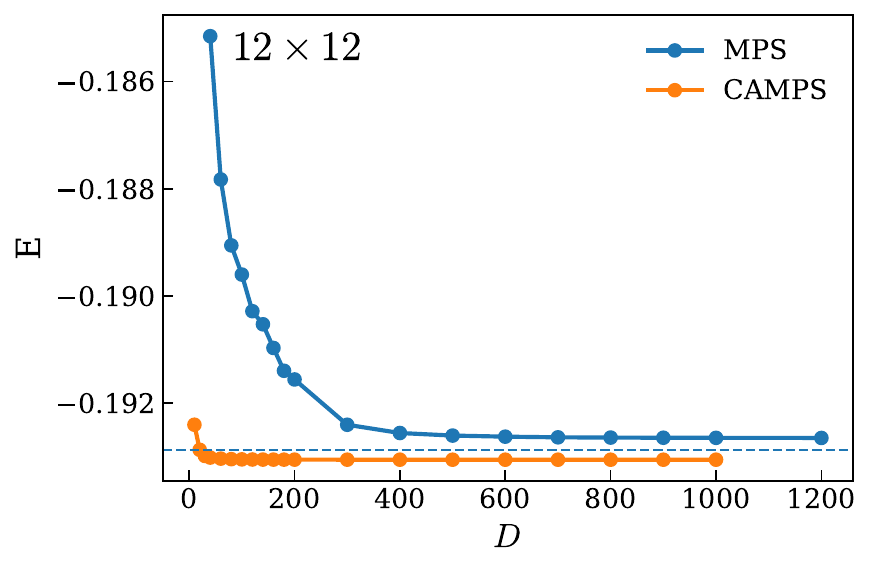}
    \includegraphics[width=80mm]{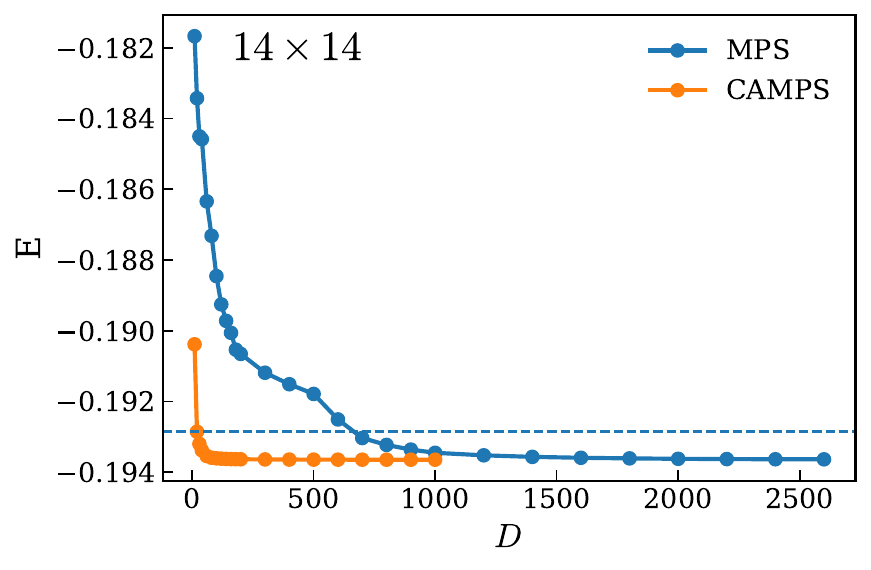}
        \caption{%The ground-state energy of the Kitaev model as a function of the bond dimension $D$ for both MPS (blue) and CAMPS (orange). The system size is $14\times14$. While the MPS energy decreases slowly and requires $D \gtrsim 1000$ to approach convergence, the CAMPS reaches a significantly lower energy already at very small bond dimensions ($D \lesssim 60$).Notably, the CAMPS result at a very small bond dimension ($D=20$) already achieves a lower variational energy than the MPS calculation with $D=600$.
        The ground-state energy of the Kitaev model as a function of the bond dimension $D$ for both MPS (blue) and CAMPS (orange) across different system sizes ($8\times8$, $10\times10$, $12\times12$, and $14\times14$). For the $14\times14$ lattice, the MPS energy decreases slowly and requires $D \gtrsim 1000$ to approach convergence, while CAMPS reaches a significantly lower energy already at very small bond dimensions ($D \lesssim 60$). Notably, the CAMPS result at a very small bond dimension ($D=20$) already achieves a lower variational energy than the MPS calculation with $D=600$. Similar trends are observed for smaller lattice sizes, with CAMPS consistently outperforming MPS in terms of energy and convergence speed.}
        \label{Kitaev result SS}
\end{figure*}

Fig.~\ref{Kitaev result SS} illustrates the ground-state energy of the Kitaev model as a function of the bond dimension $D$ for both MPS (blue) and CAMPS (orange) across various system sizes.

For the $8\times8$ lattice, where entanglement is relatively low, both MPS and CAMPS result converge with small $D$. However, CAMPS demonstrates significantly faster convergence, achieving lower energies with smaller bond dimensions.

For the $10\times10$ lattice, CAMPS continues to outperform MPS, with its results at $D=100$ surpassing those of MPS at $D=300$, highlighting its efficiency.

For the $12\times12$ lattice, MPS struggles with initialization issues, often becoming trapped in high-energy local minima. In contrast, CAMPS avoids these problems, converging rapidly to lower energy states. Remarkably, CAMPS achieves a lower energy at $D=20$ than MPS does at $D=1200$, underscoring its robustness.

For the $14\times14$ lattice, MPS convergence is slow, requiring $D \gtrsim 1000$ to approach the ground state. CAMPS, however, achieves significantly lower energies with $D \lesssim 60$, and even at $D=20$, it outperforms MPS at $D=600$.

Overall, these comparisons confirm that CAMPS systematically achieves lower variational energies at substantially reduced bond dimensions and exhibits greater optimization stability, particularly as the system size increases.

%For the $14\times14$ lattice Fig.~\ref{Kitaev result 14}, the MPS energy decreases slowly and approaches convergence only when $D \gtrsim 1000$, while CAMPS approaches convergence at $D \lesssim 60$.  The CAMPS result at $D=20$ is lower than the MPS result at $D=600$.

%For the $12\times12$ lattice Fig.~\ref{Kitaev result 12}, the MPS optimization becomes trapped in a high-energy configuration due to unfavorable random initialization, while CAMPS avoids this issue and converges rapidly. The CAMPS result at $D=20$ is lower than the MPS result at $D=1200$., illustrating its improved stability with respect to initialization and local minima.

%For the $10\times10$ result Fig.~\ref{Kitaev result 10}, the CAMPS result at $D=100$ is lower than the MPS result at $D=300$. For the $8\times8$ lattice Fig.~\ref{Kitaev result 8}, the CAMPS energy and the MPS result are similar at $D=200$. The improvement of CAMPS over MPS decreases as the system becomes smaller and relatively weakly entangled because MPS can also reach high accuracy at small bond dimensions.

%These results across multiple lattice sizes show that CAMPS achieves lower energy at substantially reduced bond dimensions and exhibits greater robustness during optimization. The method remains effective over a wide range of system sizes. The improvement becomes more significant as the computational difficulty increases.

\section{Results for different system sizes of the Kitaev Heisenberg Model}
In the main text, we presented the ground-state energy and the anti-ferromagnetic to Kitaev quantum spin liquid phase-transition point analysis for the $16 \times 16$ Kitaev--Heisenberg model. Here, we provide supplementary data for additional system sizes ($8\times8$, $10\times10$, $12\times12$, and $14\times14$) to illustrate the phase transition point results for different system sizes.

\begin{figure*}[t]
    \includegraphics[width=80mm]{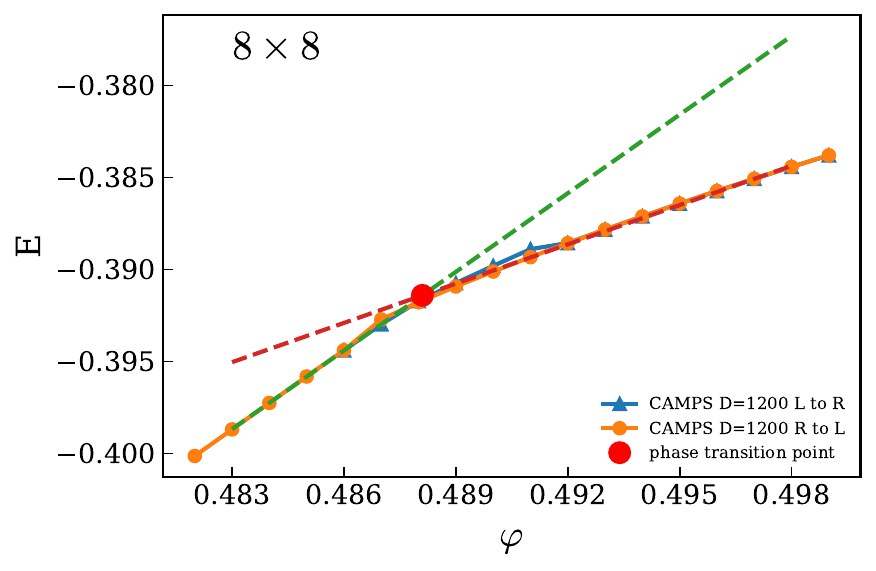}
    \includegraphics[width=80mm]{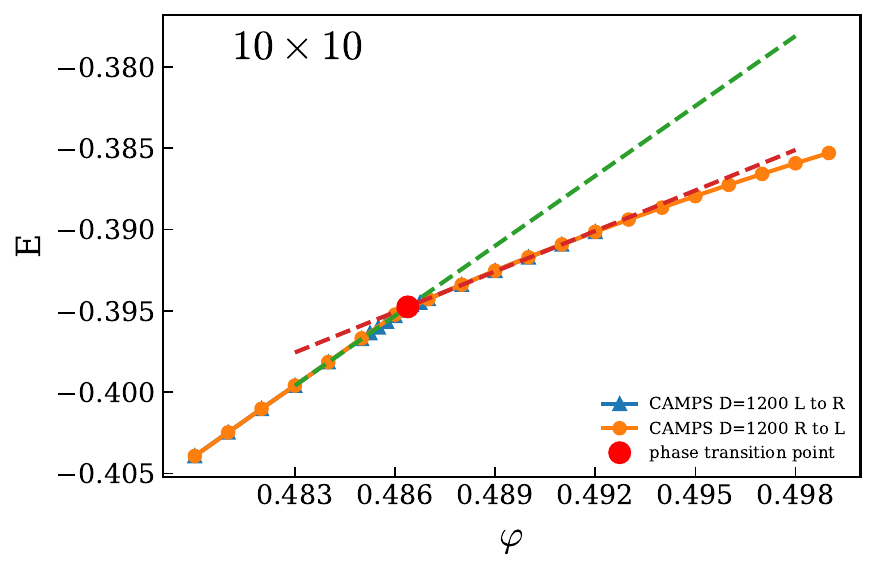}
    \includegraphics[width=80mm]{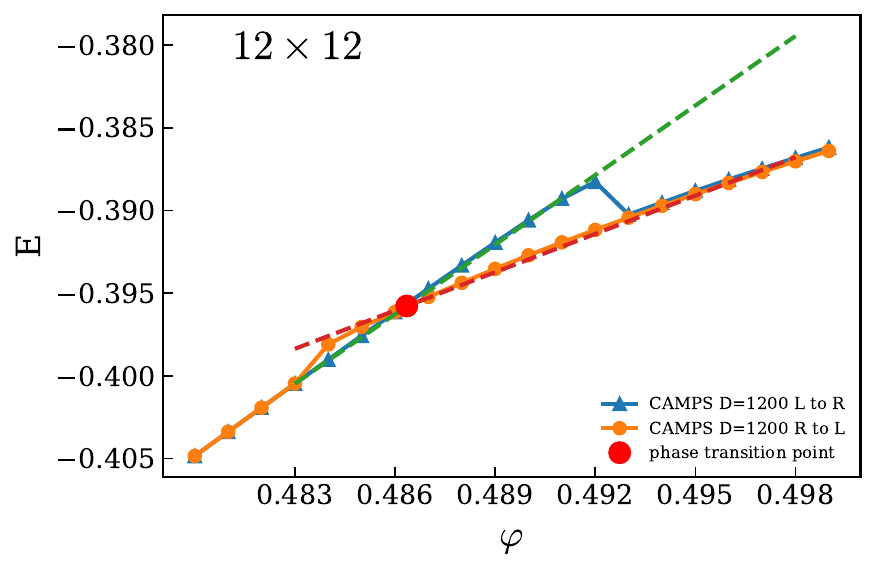}
    \includegraphics[width=80mm]{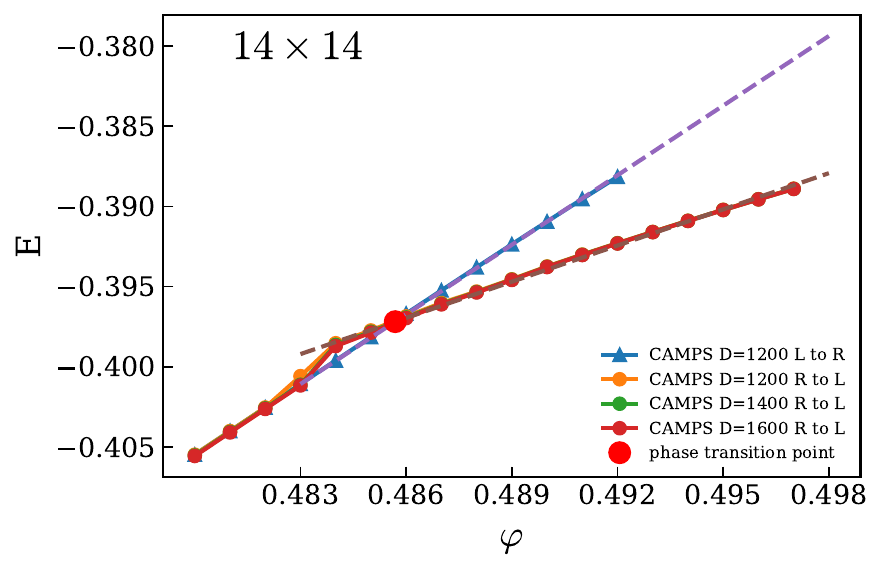}
        \caption{%Ground-state energy of the Kitaev--Heisenberg model on the $14\times14$ lattice as a function of the parameter $\varphi$. The curves labeled “L to R" indicate that each data point is initialized using the converged wavefunction from the point immediately to its left, forming a left-to-right sweep in $\varphi$. Conversely, “R to L" indicates that sweeps where each point is initialized from the converged state on its right. The dashed lines represent a linear extrapolation based on the CAMPS energies. The intersection between the two lines identifies the phase-transition point, marked by the red dot. The transition point of the $14\times14$ lattice $\varphi_c=0.48569(9)$
        Ground-state energy of the Kitaev--Heisenberg model for different system sizes as a function of the parameter $\varphi$ (in the unite of $\pi$). Since the transition is first order, we locate the critical point using an energy-crossing analysis. Specifically, we perform independent linear fits to the ground state energies on either side of the transition. The intersection of the fitted branches provides an estimate of the critical parameter $\varphi_c$, as indicated by the red dot. Specifically, we first perform simulations sweeping from left to right (increasing $\varphi$), initializing each step using the wavefunction from the previous simulation; this branch is labeled ``L to R''. Subsequently, we perform simulations sweeping from right to left (decreasing $\varphi$), similarly initializing with the wavefunction from the preceding step; this branch is labeled ``R to L''.}
        \label{Kitaev Heisenberg result}
\end{figure*}

The results are shown in Fig.~\ref{Kitaev Heisenberg result}. Because the transition is first order, we locate the critical point from an energy-crossing analysis. We perform independent linear fits to the ground-state energies on the two sides of the transition and take the intersection of the fitted branches as the estimate of $\varphi_c$ (red point in Fig.~\ref{Kitaev Heisenberg result}). We conduct simulations approaching the critical region from both directions to capture the first-order transition. Specifically, we first perform simulations sweeping from left to right (increasing $\varphi$), initializing each step using the wavefunction from the previous simulation; this branch is labeled ``L to R'' in Fig.~\ref{Kitaev Heisenberg result}. Subsequently, we perform simulations sweeping from right to left (decreasing $\varphi$), similarly initializing with the wavefunction from the preceding step; this branch is labeled ``R to L'' in Fig.~\ref{Kitaev Heisenberg result}.

%%%%%%%%%%%%Refs
% \bibliography{main}
%%%%%%%%%%%%